\documentstyle[twoside,aps]{revtex}
\newcommand{\snl}{\stackrel{(0)}{<}}
\newcommand{\snr}{\stackrel{(0)}{>}}
\begin{document}
\draft
\title{Rayleigh-Schr\"{o}dinger Perturbation Theory \\ Based on Gaussian
       Wavefunctional Approch}
\author{Wen-Fa Lu$^{a,b,c}$, Chul Koo Kim$^{a,c}$, Jae Hyung Yee$^{a}$
        and Kyun Nahm$^d$}
\address{$^a$ Institute of Physics and Applied Physics, Yonsei University,
Seoul 120-749, Korea \\
 $^b$ Department of Applied Physics, Shanghai Jiao Tong University,
Shanghai 200030, the People's Republic of China
   \thanks{permanent address,E-mail: wenfalu@online.sh.cn} \\
 $^c$ Center for Strongly Correlated Materials Research,
Seoul National University, Seoul 151-742, Korea \\
 $^d$ Department of Physics, Yonsei University, Wonju 220-710, Korea
    }
\maketitle

\begin{abstract}
A Rayleigh-Schr\"{o}dinger perturbation theory based on the Gaussian
wavefunctional is constructed. The method can be used for calculating the
energies of both the vacuum and the excited states. A model calculation is
carried out for the vacuum state of the $\lambda\phi^4$ field theory.
\end{abstract}
\vspace{24pt}
PACS numbers : 11.10.-z; 11.10.Lm; 11.15.Tk .

\section{Introduction}
\label{1}

The Gaussian wave-functional approach (GWFA) \cite{1}, or Gaussian
approximation, has become a powerful and important tool to extract
non-perturbative result of quantum field theory \cite{2,3,4}, finite
temperature field theory \cite{5}, and condensed matter systems \cite{6,7}
since Stevenson's advocation of the method about two decades ago \cite{8}.
The Gaussian effective potential (GEP) obtained from the GWFA provides a good
starting point for further investigations of various systems
\cite{9,10,11,12,13,14,15,16,17}. Moreover, this approximation can also be
used for realizing some novel ideas \cite{18}. However, the GWFA is
essentially a variational approximation, and hence improvement of the
obtained result is not straightorward. So far, there exist mainly two methods
to improve the GWFA. One way is to continue using the variational method with
more elaborate, non-Gaussian trial wavefunctionals. For example, K\"ummel and
his collabrators developed the coupled cluster method and obtained results
beyond the Gaussian approximation of the $\lambda\phi^4$ and $\phi^6$ models
\cite{9}; Ritschel and his collaborators constructed a non-Gaussian trial
wave functionals through a nonlinear canonical transformation \cite{10}; In
order to investigate ($3+1$)-dimentional $\lambda\phi^4$ field theory,
Yotsuyanagi proposed an improved scheme of the GWFA by adopting a BCS-type
wavefunctional \cite{11}. Another way of improveing the GWFA is to use
appropriate expansions which give the GEP in their lowest order. In this
aspect, Okopi\'{n}ska developed an optimized expansion method to calculate
the generating functional with the Euclidean formalism \cite{12} ; In the
same Euclidean formalism, Stancu and Stevenson formulated a slightly
different expansion scheme and calculated the post-GEP in the spirit of the
background-field method \cite{13}; Based on the GEP, Cea proposed a
generalized GEP with a variational basis and carried out the calculation with
the help of the standard perturbation technique in quantum field theory
\cite{14}; In the late 1990s, within the Minkowski formalism, one of the
authors (Yee) and his collabrators developed the background field method to
give an expansion of the effective action around the Gaussian approximate
results \cite{15}. Recently, in order to calculate the partition function of
a fermionic system, two of the authors (Kim and Nahm) and their collaborator
proposed a variational perturbation scheme based on the functional integral
without resorting to the background field method \cite{16}. Additionally,
Solovtsov $et \ al$. proposed a kind of variational perturbation theory to
calculate the effective potential \cite{17}. All these investigations improve
the GEP or Gaussian-approximate result with miscellaneous degrees of success.
Nevertheless, further investigations are needed still to achieve better
approximations.

We note that Rayleigh-Schr\"{o}dinger perturbation theory is one of the basic
techniques in nonrelativistic quantum mechanics \cite{19}. It is also widely
used in statistical mechanics \cite{20} and, recently, generalized to quantum
field theory \cite{21}. Conventionally, Rayleigh-Schr\"{o}dinger perturbation
theory is based on an exactly solvable part of the Hamiltonian of a system.
Recently, two of the authors (Kim and Nahm) and their collabarators
calculated the energies of an anharmonic oscillator by combining
Rayleigh-Schr\"{o}dinger perturbation theory and variational method and
obtained satisfactory results \cite{22}. In this paper, we develop a
variational Rayleigh-Schr\"{o}dinger perturbation theory based on the
Gaussian wave-functional approach (RSPTGA) in quantum field theory. Different
from the schemes mentioned in the last paragraph, RSPTGA can be used to
calculate the energies of both the vacuum and excited states. Applying RSPTGA
to $\lambda\phi^4$ model, we calculate the vacuum state energy up to the
third order. The result improves the GEP substantially and indicates a fast
convergence. Comparing with the result in Ref.~\cite{14}, our result up to
second order is shown to introduce an additional term.

Next section, we construct a quasi-free-field eigenstate set based on the
GWFA, and develop RSPTGA based on the eigenset. In Sect. III, the vacuum
state energy of the $\lambda\phi^4$ field theory will be calculated using
RSPTGA. Finally, physical implication of the present results will be
discussed.

\section{Rayleigh-Schr\"{o}dinger perturbation theory Based on the GWFA}
\label{2}

In this section, we briefly introduce the GWFA, and construct a
quasi-free-field eigenstate set. Based on this eigenstate set, we construct
RSPTGA.

We consider a model with the Lagrangian density
\begin{equation}
{\cal L}={\frac {1}{2}}\partial_\mu \phi_x \partial^\mu \phi_x
-V(\phi_x) \;,
\end{equation}
where $x=(x^1,x^2, \cdots, x^D)$ represents a position in $D$-dimensional
space, $\phi_x\equiv \phi(\vec{x})$ is the field at $x$, and the potential
$V(\phi_x)$ has a Fourier reprensentation in a sense of tempered
distributions \cite{23}. Many model potentials, such as polynomial models,
sine-Gordon and sinh-Gordon models, possess this property. According to
Ref.~\cite{4} (JPA), the GWFA produces the best trial vacuum wavefunctional
\begin{equation}
|0\snr = {\cal N} \exp\{ -{\frac {1}{2}}\int_{x,y}
            (\phi_x - \varphi)f_{xy}(\phi_y - \varphi)\}
\end{equation}
where, ${\cal N}$ is the normalization constant ($i.e., \snl 0|0\snr=1$),
$\int_{x,y}\equiv \int d^D x d^D y$, and $f_{xy}=\int d^D p f(p)e^{i p(x-y)}$
with $p=(p^1, p^2, \cdots, p^D)$. The classical constant $\varphi$ is equal
to the Gaussian-vacuum expectation value of $\phi_x$,
\begin{equation}
\varphi=\snl 0|\phi_x|0\snr \;.
\end{equation}
Here,
\begin{equation}
f(p)=\sqrt{p^2+\mu^2(\varphi)}  \;,
\end{equation}
and
\begin{equation}
\mu^2(\varphi)=\int^\infty_{-\infty} {\frac {d\alpha}{2\sqrt{\pi}}}e^{-{\frac
 {\alpha^2}{4}}} V^{(2)}({\frac {\alpha}{2}}\sqrt{I_1 (\mu(\varphi))}
 +\varphi) \;,
\end{equation}
where $V^{(n)}(z)={\frac {d^nV(z)}{dz^n}}=
      \int^\infty_{-\infty}{\frac {dq}{\sqrt{2\pi}}}(iq)^n
          {\tilde{V}}(q)e^{iq z}$
(${\tilde{V}}(q)$ is the Fourier representation of $V(z)$)
and $I_n(Q)=\int {\frac {d^D p}
{(2\pi)^D}}{\frac {\sqrt{p^2+Q^2}}{(p^2+Q^2)^n}}$. The GEP of the system,
Eq.(1), is given by
\begin{equation}
{\cal V}_G (\varphi)= {\frac {1}{2}}I_0(\mu)-{\frac {\mu^2}{4}}I_1 (\mu)
+\int^\infty_{-\infty} {\frac {d\alpha}{2\sqrt{\pi}}}e^{-{\frac
     {\alpha^2}{4}}} V({\frac {\alpha}{2}}\sqrt{I_1(\mu)}+\varphi)
\end{equation}
with $\mu$ chosen from the three possibilities: solution of Eq.(5), $\mu=0$
and $\mu \to \infty$ \cite{4}(JPG) (Hereafter, we will write $\mu(\varphi)$
as $\mu$ for simplicity except for special cases). We note that when
${\cal V}_G (\varphi)$ has the absolute minimum at $\varphi_0$,
$\mu(\varphi_0)$ becomes just the physical mass and ${\cal V}_G (\varphi_0)$
represents the vacuum state energy. The symmetry of the vacuum can be
discussed using Eqs.(3)$-$(6).

For the Gaussian vacuum, Eq.(2), one can construct the following annihilation
and creation operators \cite{4} (ZPC):
\begin{equation}
A_f(p) = ({\frac {1}{2(2\pi)^D f(p)}})^{1/2}\int_x e^{-ipx}
   [f(p)(\phi_x - \varphi) + i\Pi_x]
\end{equation}
and
\begin{equation}
A^\dagger_f(p) = ({\frac {1}{2(2\pi)^D f(p)}})^{1/2}\int_x e^{ipx}
   [f(p)(\phi_x - \varphi) - i\Pi_x]
\end{equation}
with the relations $[A_f(p),A^\dagger_f(p')]=\delta (p'-p)$ and
$A_f(p)|0\snr=0$. Here, $\Pi_x\equiv -i{\frac {\delta}{\delta \phi_x}}$
is the canonical conjugate operator to $\phi_x$ with the commutation relation,
$[\phi_x,\Pi_y]=i\delta(x-y)$. Based on these operators, one can construct
the quasi-free-field Hamiltonian,
\begin{equation}
H_0=\int dp f(p) A^\dagger_f(p)A_f(p)=\int_x [ {\frac {1}{2}}\Pi^{2}_x
                +{\frac {1}{2}}(\partial_x \phi_x)^2 +
                {\frac {1}{2}}\mu^2 (\phi_x-\varphi)^2
                -{\frac {1}{2}}I_0(\mu)]    \;.
\end{equation}
The Gaussian vacuum, Eq.(2), is the ground state of $H_0$ with zero energy
eigenvalue $E_0^{(0)}$. The excited states of $H_0$ are \cite{21}
\begin{equation}
|n\snr ={\frac {1}{\sqrt{n!}}} \prod_{i=1}^n A^\dagger_f(p_i)|0\snr
          ,  \; \; n= 1, 2, \cdots, \infty
\end{equation}
with the corresponding energy eigenvalue
\begin{equation}
E^{(0)}_n=\sum_{i=1}^n f(p_i) .
\end{equation}
Note that Eq.(2) is not a naive vacuum, since it contains information on the
interacting system, Eq.(1). Obviously, the wavefunctionals $|n\snr$ and
$|0\snr$ are orthogonal and normalized to $\snl n|n\snr ={\frac {1}{n!}}
\sum_{P_i(n)}\prod_{k=1}^n\delta(p'_{k}-p_{i_k})$ (here $P_i(n)$ represents a
permutation of the set $\{i_k\}=\{1,2,\cdots,n\}$ and the summation is over
all $P_i(n)s$). $|n\snr$ describes a $n$-particle state with the
continuous momenta $p_1, p_2,\cdots,p_n$. $|0\snr$ and $|n\snr$ with $n=1, 2,
\cdots, \infty$ constitute the complete set for $H_0$, which we call
quasi-free-field complete set.

Based on the above complete set, one can readily apply the conventional
Rayleigh-Sch\"{o}dinger perturbation technique to calculate the energy of the
system, Eq.(1). In order to do so, we write the Hamiltonian of the system
Eq.(1) as $H=H_0+H_I=H_0+(H-H_0)$ with
\begin{equation}
H_I=\int_x [-{\frac {1}{2}}\mu^2 (\phi_x-\varphi)^2 +{\frac {1}{2}}I_0(\mu)
      +V(\phi_x)] \;.
\end{equation}
Following the Rayleigh-Sch\"{o}dinger perturbation procedure \cite{19,21},
one can obtain the wavefunctionals and energies for the vacuum and excited
states respectively,
\begin{equation}
|n>=\sum_{l=0}^{\infty} [Q_n {\frac {1}{H_0-E_n^{(0)}}}(E_n-E_n^{(0)}-H_I)]^l
      |n\snr       \;,
\end{equation}
and 
\begin{equation}
E_n(\varphi)=E_n^{(0)}+\sum_{l=0}^{\infty} \snl n| H_I
        [Q_n {\frac {1}{H_0-E_n^{(0)}}}(E_n-E_n^{(0)}-H_I)]^l
      |n\snr
\end{equation}
with $Q_n=\sum_{j\not=n}^\infty \int d^D p_1d^D p_2\cdots d^D p_j|j\snr
\snl j|$.

For the case of $n=0$ and $l=0$, Eq.(14) gives the vacuum enegy up
to the first order $E_0^{1}=E_0^{(0)}+E_0^{(1)}=\int_x {\cal V}_G (\varphi)$
which is just the product of the GEP and the space volume. Thus, employing
Eq.(14) with $n=0$, one can get the effective potential for the system,
Eq.(1),
\begin{equation}
{\cal V}_{RS} (\Phi) = {\frac {E_0(\varphi)}{\int d^D x}}
\end{equation}
with
\begin{equation}
\Phi={\frac {<0|\phi_x|0>}{<0|0>}}       \;,
\end{equation}
which takes the GEP, Eq.(6), as the first-order approximation. Eq.(16)
implies that $\Phi$ should replace $\varphi$ in the calculation of the
effective potential. It is evident that $\varphi$ in Eq.(3) is the zeroth
order approximation, $\Phi^{(0)}$ of $\Phi$. When Eq.(15) is truncated at
$n$th order, one should also truncate Eq.(16) at the same order \cite{12,13}.

In the case of $n\not=0$, Eq.(14) is the excited-state effective potential
of the system \cite{8}. Additionally, Eq.(14) becomes the vacuum and
excited-state energies for the symmetric phase of the system if $\varphi=0$
is chosen in the scheme. In the next section, we apply RSPTGA to the vacuum
state of the $\lambda\phi^4$ field theory.

\section{Application to $\lambda\phi^4$ field theory}
\label{3}

In this section, we consider the potential,
$V(\phi_x)={\frac {1}{2}}m^2\phi_x^2+{\frac {\lambda}{4!}}\phi_x^4$, which
was widely studied in connection with the Gaussian approximation \cite{14}.
For this system, Eq.(5) gives rise to
\begin{equation}
\mu^2=m^2+{\frac {1}{2}}\lambda\varphi^2 +{\frac {1}{4}}\lambda I_1(\mu) \;,
\end{equation}
and from Eq.(6), one readily obtains
\begin{equation}
{\cal V}_G (\varphi)={\frac {1}{2}}m^2\varphi^2+{\frac {1}{4!}}\lambda
     \varphi^4
   +{\frac {1}{2}}I_0(\mu) -{\frac {1}{32}}\lambda I_1^2(\mu) \;,
\end{equation}
which is just Eq.(2.22) in Ref.~\cite{14}(PRD) \footnote{our
notation $I_n(\mu)$ is different from Eq.(2.21) in Ref.~\cite{14}(PRD).}.
To obtain the effective potential of the $\lambda\phi^4$ field theory up to
a given order according to Eq.(15), we first calculate the following matrix
elements of $H_I$ :
\begin{eqnarray}
\snl n|H_I|n\snr &=&\snl 0|H_I|0\snr \snl n|n\snr + {\frac {1}{n!}}{\frac
    {\lambda}{4(2\pi)^D}}\sum_{\{P_{ij}(n-2)\}}
     \prod_{k=1}^{n-2}\Delta(p_{i_k}-p'_{j_k}) \nonumber  \\
     && \cdot  \delta(p'_{j_{(n-1)}}+p'_{j_n}
      -p_{i_{(n-1)}}-p_{i_n})
       [f(p'_{j_{(n-1)}})f(p'_{j_n})
      f(p_{i_{(n-1)}})f(p_{i_n})]^{-{\frac {1}{2}}} \;,
\end{eqnarray}
\begin{eqnarray}
\snl n|H_I|n-1\snr &=&{\frac {1}{\sqrt{n!(n-1)!}}}\biggl\{
   {\frac {\lambda\varphi}{2\sqrt{2(2\pi)^D}}}
     \sum_{P_{ij}(n-2)}\prod_{k=1}^{n-2}
      \Delta(p_{i_k}-p'_{j_k})  \nonumber  \\ && \cdot\delta(p'_{j_{(n-1)}}
      -p_{i_{(n-1)}}-p_{i_n})[f(p'_{j_{(n-1)}})
      f(p_{i_{(n-1)}})f(p_{i_n})]^{-{\frac {1}{2}}}   \nonumber   \\
      &&  +\sqrt{\frac {(2\pi)^D}{2}}
         (\mu^2-{\frac {\lambda}{3}}\varphi^2)\varphi
         \sum_{P_{ij}(n-1)}\prod_{k=1}^{n-1}
      \Delta(p_{i_k}-p'_{j_k})\delta(p_{i_n})
      [f(p_{i_n})]^{-{\frac {1}{2}}} \biggr\} \;,
\end{eqnarray}
\begin{eqnarray}
\snl n|H_I|n-2\snr &=&
   {\frac {1}{\sqrt{n!(n-2)!}}}{\frac {\lambda}{4(2\pi)^D}}
     \sum_{P_{ij}(n-3)}\prod_{k=1}^{n-3}\Delta(p_{i_k}-p'_{j_k})
     \nonumber  \\
   && \cdot\delta(p'_{j_{(n-2)}}-p_{i_{n-2}}
      -p_{i_{(n-1)}}-p_{i_n})
       [f(p'_{j_{(n-2)}})f(p_{i_{n-2}})
      f(p_{i_{(n-1)}})f(p_{i_n})]^{-{\frac {1}{2}}}  \; ,
\end{eqnarray}
\begin{eqnarray}
\snl n|H_I|n-3\snr &=&
   {\frac {1}{\sqrt{n!(n-3)!}}}{\frac {\lambda\varphi}{2\sqrt{2(2\pi)^D}}}
     \sum_{P_{ij}(n-3)}\prod_{k=1}^{n-3}\Delta(p_{i_k}-p'_{j_k})
      \nonumber  \\  && \cdot\delta(p_{2_{(n-2)}}
      +p_{i_{(n-1)}}+p_{i_n}) [f(p_{i_{(n-2)}})
      f(p_{i_{(n-1)}})f(p_{i_n})]^{-{\frac {1}{2}}}
\end{eqnarray}
and
\begin{eqnarray}
\snl n|H_I|n-4\snr &=&
   {\frac {1}{\sqrt{n!(n-4)!}}}{\frac {\lambda}{4(2\pi)^D}}
     \sum_{P_{ij}(n-4)}\prod_{k=1}^{n-4}\Delta(p_{i_k}-p'_{j_k})
    \nonumber  \\  && \cdot\delta(p_{i_{(n-3)}}+p_{i_{n-2}}
      +p_{i_{(n-1)}}+p_{i_n})
      [f(p_{i_{(n-3)}})f(p_{i_{(n-2)}})
      f(p_{i_{(n-1)}})f(p_{i_n})]^{-{\frac {1}{2}}}  \;,
\end{eqnarray}
with
\[
   \Delta(p_{i_k}-p'_{j_k})=\left \{
    \begin{array}{ll}
     0,& \ \ \ \ \  {\rm  for} \ \ \ \ \ k<0 \\
     1,& \ \ \ \ \  {\rm  for} \ \ \ \ \ k=0 \\
     \delta(p_{i_k}-p'_{j_k}), & \ \ \ \ \  {\rm  for} \ \ \ \ \ k>0
    \end{array} \;.  \right.
\]
Here, the index $i_k\in \{1,2,\cdots,n\}$ with $k=1,2,\cdots,n$ corresponds
to $|n\snr$ and $j_k\in\{1,2,\cdots,n'\}$ with $k=1,2,\cdots, n'$ to
$\snl n'|$. $P_{ij}(l)$ represents a given permutation of $l$ momenta
$p_{i_1},p_{i_2},\cdots,p_{i_l}$ paired respectively with $p'_{j_1},p'_{j_2},
\cdots,p'_{j_l}$, and $\sum_{P_{ij}(l)}$ is over all different $P_{ij}(l)s$.
For any $P_{ij}(l)$, $i_1, i_2, \cdots, i_l$ are different from one another,
and so are $j_1, j_2, \cdots, j_l$. Employing the above matrix elements, a
straightforward, yet lengthy calculation according to Eq.(15) gives the
effective potential of the $\lambda\phi^4$ field theory up to the third order
as
\begin{eqnarray}
{\cal V}^{iii} (\varphi) \equiv {\frac {E_0^{iii}}{(2\pi)^D\delta(0)}}
    &=&{\cal V}_G (\varphi)   \nonumber  \\
     &&-{\frac {1}{2}}{\frac {1}{\mu^2}}\varphi^2
    (\mu^2-{\frac {\lambda}{3}}\varphi^2)^2
      -{\frac {A}{48}}{\frac {\lambda^2}{\mu^2}}\varphi^2
       -{\frac {B}{384}}{\frac {\lambda^2}{\mu^2}}   \nonumber  \\
     && +{\frac {A+A_1+A_2}{48}}{\frac {\lambda^2}{\mu^4}}\varphi^2
           (\mu^2-{\frac {\lambda}{3}}\varphi^2)
           +{\frac {2 B_1+B_2}{128}}{\frac {\lambda^3}{\mu^4}}\varphi^2
           +{\frac {C}{512}}{\frac {\lambda^3}{\mu^4}}
\end{eqnarray}
with
\[
 \begin{array}{l}
  A=\int {\frac {dx}{(2\pi)^D}} {\frac {dy}{(2\pi)^D}}
   [f_1(x)f_1(y)f_1(x+y)]^{-1}[f_1(x)+f_1(y)+f_1(x+y)]^{-1} \;, \\
  A_1=\int {\frac {dx}{(2\pi)^D}} {\frac {dy}{(2\pi)^D}}
   [f_1(x)f_1(y)f_1(x+y)]^{-1}[1+f_1(x)+f_1(y)+f_1(x+y)]^{-1}  \; ,  \\
  A_2=\int {\frac {dx}{(2\pi)^D}} {\frac {dy}{(2\pi)^D}}
   [f_1(x)f_1(y)f_1(x+y)]^{-1}[f_1(x)+f_1(y)+f_1(x+y)]^{-1} \\
   \hspace{1cm}  \cdot  [1+f_1(x)+f_1(y)+f_1(x+y)]^{-1}  \; ,  \\
  B=\int {\frac {dx}{(2\pi)^D}} {\frac {dy}{(2\pi)^D}} {\frac {dz}{(2\pi)^D}}
   [f_1(x)f_1(y)f_1(z)f_1(x+y+z)]^{-1}  \\
   \hspace{1cm} \cdot [f_1(x)+f_1(y)+f_1(z)+f_1(x+y+z)]^{-1} \; , \\
  B_1=\int {\frac {dx}{(2\pi)^D}}{\frac {dy}{(2\pi)^D}}
      {\frac {dz}{(2\pi)^D}}
   [f_1(x)f_1(y)f_1(z)f_1(x+y)f_1(x+y+z)]^{-1}  \\
   \hspace{1cm}  \cdot  [f_1(x)+f_1(y)+f_1(x+y)]^{-1}
   [f_1(x)+f_1(y)+f_1(z)+f_1(x+y+z)]^{-1}  \\
  B_2=\int {\frac {dx}{(2\pi)^D}}{\frac {dy}{(2\pi)^D}}
      {\frac {dz}{(2\pi)^D}}
   [f_1(x)f_1(y)f_1(z)f_1(x+y)f_1(x+z)]^{-1}  \\
   \hspace{1cm}  \cdot  [f_1(x)+f_1(y)+f_1(x+y)]^{-1}
   [f_1(x)+f_1(z)+f_1(x+z)]^{-1}  \\
  C=\int {\frac {dx}{(2\pi)^D}}{\frac {dy}{(2\pi)^D}}
      {\frac {dz}{(2\pi)^D}} {\frac {d\omega}{(2\pi)^D}}
   [f_1(x)f_1(y)f_1(z)f_1(\omega)f_1(x+y+z)f_1(x+y+\omega)]^{-1}  \\
   \hspace{1cm}  \cdot  [f_1(x)+f_1(y)+f_1(z)+f_1(x+y+z)]^{-1}
   [f_1(x)+f_1(y)+f_1(\omega)+f_1(x+y+\omega)]^{-1}
  \end{array}
\]
and $f_1(w)=\sqrt{1+w^2}$. Here, we take $\varphi$ as the zeroth order
approximation of $\Phi$. In Eq.(24), the second line is the second order
correction and the third line the third order correction to the GEP.

From Eq.(24), one can see that after ${\cal V}_G(\varphi)$ is renormalized,
its corrections will be finite and, accordingly, a further renormalization
procedure is not needed for higer order corrections. In the case of (0+1)
dimensions, our second-order result with $\varphi=0$ is consistent with
Eq.(14) in Ref.~\cite{22}. Furthermore, numerical calculation for the case of
(1+1) dimensions indicates that the effective potential with the second-order
correction predicts existence of a second-order phase transition. We also
note that the second-order correction improves GEP substantially, and the
third order correction is vanishingly small. This can be seen by comparing
the coeficients of the first, second and third terms in the second line of
Eq.(24) with corresponding ones in the third line, respectively. This
indicates that RSPTGA has a fast convergence in general.

\section{Discussion and Conclusion}
\label{5}

In this paper, a Rayleigh-Schr\"{o}dinger perturbation theory based on the
GWFA within the framework of quantum field theory is proposed. Since the
theory is based on the Gaussian approximation, it provides a systematical
tool for controlling the Gaussian approximation. It can be used not only for
calculating the effective potential but also for considering excited states.
When one is interested in symmetric phase, the vacuum and excited-state
energies can be calculated beyond the Gaussian approximation by RSPTGA using
$\varphi=0$. Application of RSPTGA to the $\lambda\phi^4$ field theory shows
that it can improve the GEP substantially with a fast convergence.

We note that RSPTGA predicts existence of a second-order phase transition in
the (1+1)-dimensional $\lambda\phi^4$ field theory, although the critical
coupling is very small. It may be attributed to the fact that for the
second-order case we approximated $\Phi=<0|\phi_x|0>\bigl| _{l=2}$ as
$\varphi$. We also note that the second-order result of Eq.(24) has an
additional term $-{\frac {1}{2}}{\frac {1}{\mu^2}}\varphi^2
(\mu^2-{\frac {\lambda}{3}}\varphi^2)^2$ which does not appear in
Ref.~\cite{14}. In fact, Eq.(12) gives, for the $\lambda\phi^4$ field theory,
\begin{equation}
H_I=\int_x\{{\cal V}_G (\varphi) +\varphi(\mu^2-{\frac {\lambda}{3}}\varphi^2)
    :(\phi_x-\varphi):+{\frac {\lambda}{3!}}\varphi
    :(\phi_x-\varphi)^3:+{\frac {\lambda}{4!}}:(\phi_x-\varphi)^4:\} ,
\end{equation}
which is equal to Eq.(4.19) in Ref.~\cite{14} (PRD) except the
$\int_x {\cal V}_G (\varphi)$ term, where the colons mean normal ordering
with respect to the Gaussian vacuum, Eq.(2). The additional term in Eq.(24)
arises from the linear term $\varphi(\mu^2-{\frac {\lambda}{3}}\varphi^2)
:(\phi_x-\varphi):$, and disappears when the constraint Eq.(5.18) in
Ref.~\cite{14} (PRD) is adopted in our scheme; that is, when $\Phi$ in
Eq.(16) is taken as $\varphi$. However, we note that $\varphi$ is simply the
Gaussian-vacuum average value of the field operator (see Eq.(3)) and
$\Phi$ is not equal to $\varphi$ when higher order contributions are
calculated \cite{10,12,13}. Moreover, the variational procedure which led to
the GEP, Eq.(18) produces the extremum condition, Eq.(17) ($i.e.$, Eq.(2.19)
in Ref.~\cite{14} (PRD)), and Eq.(5.18) in Ref.~\cite{14} (PRD) at any
truncated order is not compatible with this extremum condition.

In closing the paper, we like to point out that it is straightforward to
generalize RSPTGA to other cases, such as a Fermion field system. In fact, a
Rayleigh-Schr\"{o}dinger perturbation technique based on the variational
results has been applied to a polaron problem \cite{23}. Moreover, one of the
authors (Lu) developed the GWFA in thermofield dynamics \cite{25}\cite{5}(Lu).
Based on it, it is possible to develop RSPTGA within the framework of
thermofield dynamics which will be useful for treating finite temperature
cases. Finally, instead of performing the variational procedure in the GWFA
as described above, the extremization process with respect to $\mu$ can be
carried out after truncating the series of Eq.(14) at some given order
\cite{12}. This procedure will lead to a slightly different variaion of
RSPTGA \cite{8}. It may have its own advantages or peculiarities over the
scheme developed here.

\acknowledgments
This project was supported by the Korea Research Foundation
(99$-$005$-$D00011). Lu's work was also supported in part by the National
Natural Science Foundation of China under the grant No. 19875034.

\end{document}